\documentclass[]{article}

\title{Radiation Like Scalar Field and Gauge Fields in Cosmology for a theory with Dynamical Time}
\author{David Benisty and E.I. Guendelman 
\\{\small Department of Physics, Ben Gurion University of the Negev} \\{\small P.O.Box 653, IL-84105Beer-Sheva, Israel}}

\begin{document}
  \maketitle

\begin{abstract}
Cosmological solutions with a scalar field behaving as radiation are obtained, in the context of gravitational theory with dynamical time. The solution requires the spacial curvature of the universe k, to be zero, unlike the standard radiation solutions, which do not impose any constraint on the spacial curvature of the universe. This is because only such $ k=0 $ radiation solutions poses a homothetic Killimg vector. This kind of theory can be used to generalize electromagnetism and other gauge theories, in curved space time, and there are no deviations from standard gauge filed equation (like Maxwell equations) in the case there exist a conformal Killing vector. But there could be departures from Maxwell and Yang Mills equations, for more general space times. 
\end{abstract}

\section{Introduction}
There have been theoretical approches to gravity theories, where a fundamental constraint is implemented. For example in the Two Measures Theories [1] one works, in addition to the regular measure of integration in the action $ \sqrt{-g} $, also with another measure which is also a denstiy and which is also a total derivative. In this case, one can use for constructing this measure 4 scalar fields $ \varphi_{a} $, where $ a=1,2,3,4 $. Then we can define the density $ \Phi=\varepsilon^{\alpha\beta\gamma\delta}\varepsilon_{abcd}\partial_{\alpha}\varphi_{a}\partial_{\beta}\varphi_{b}\partial_{\gamma}\varphi_{c}\partial_{\delta}\varphi_{d} $, and then we can write an action that uses both of these densities:
 
\begin{equation}
S=\int d^{4}x\Phi\mathcal{L}_{1}+\int d^{4}x\sqrt{-g}\mathcal{L}_{2}
\end{equation}
As a consequence of the variation with respect to the scalar fileds $ \varphi_{a} $, assuming that $ \mathcal{L}_{1} $
 and $ \mathcal{L}_{2} $
  are independed of the scalar fileds $\varphi_{a} $, we obtain that:

\begin{equation}
A_{a}^{\alpha}\partial_{\alpha}\mathcal{L}_{1}=0
\end{equation}
 
 where $ A_{a}^{\alpha}=\varepsilon^{\alpha\beta\gamma\delta}\varepsilon_{abcd}\partial_{\beta}\varphi_{b}\partial_{\gamma}\varphi_{c}\partial_{\delta}\varphi_{d} $. Since $ det[A_{a}^{\alpha}]\sim\Phi^{3} $ as one easily see, then that for $ \Phi\neq0 $ ,(2) implies that $ \mathcal{L}_{1}=M=Const $. This result can expresed as a covariant conservation of a stress energy momentum of the form $ T_{\left(\Phi\right)}^{\mu\nu}=\mathcal{L}_{1}g^{\mu\nu} $, and using the 2nd order formalism, where the covariant derivative of $ g_{\mu\nu} $ is zero, we obtain that $ \nabla_{\mu}T_{\left(\Phi\right)}^{\mu\nu}=0 $, implies ${\partial_{\alpha}\mathcal{L}_{1}=0}$.
   This suggest generalizing the idea of the Two Measures Theory, by imposing the covariant conservation of a more nontrivial kind of energy momenetum tensors, which we denote as $ T_{\left(\chi\right)}^{\mu\nu} $ [2].
   Therfore we consider an action of the form:
\begin{equation}
S=\mathcal{S}_{\left(\chi\right)}+\mathcal{S}_{\left(R\right)}=\int d^{4}x\sqrt{-g}\chi_{\mu;\nu}T_{\left(\chi\right)}^{\mu\nu}+\frac{1}{16\pi G}\int d^{4}x\sqrt{-g}R
\end{equation}
   
   where $ \chi_{\mu;\nu}=\partial_{\nu}\chi_{\mu}-\Gamma_{\mu\nu}^{\lambda}\chi_{\lambda} $. If we assume $ T_{\left(\chi\right)}^{\mu\nu} $ to be independent of$  \chi_{\mu} $
     and having $ \Gamma_{\mu\nu}^{\lambda} $ being defined as the Christoffel Connection Coefficients, then the variation with respect to $ \chi_{\mu} $ gives a covariant conservation $ \nabla_{\mu}T_{\left(\chi\right)}^{\mu\nu}=0 $. Notice the fact that the energy density is the canonically conjugated variable to $ \chi^{0} $, which is what we expect from a dynamical time (here represented by index of time).

\section{Solutions with homothetic vector filed}
Interesting particular cases of cosmological solutions are obtained when the vector filed $ \chi^{\mu} $  has the property of being a homothetic vector filed:

\begin{equation}
\mathcal{L}_{\chi} g_{\mu\nu} \equiv \chi_{\mu;\nu}+\chi_{\nu;\mu}=\frac{\chi}{2}g_{\mu\nu}
\end{equation}

when $ \chi=\chi_{;\lambda}^{\lambda} $ is a constant trace of $ \chi_{\mu;\nu} $ (for homothetic Killing vector). \\The effective stress energy momentum tensor, from variation of the metric using the action (3) (using the identities in appendix A) is:

\begin{equation}
 T_{eff}^{\mu\nu}=\frac{\chi T(\chi)}{8}g^{\mu\nu}+\frac{\chi}{4}g_{\alpha\beta}\frac{\partial}{\partial g_{\mu\nu}}T_{(\chi)}^{\alpha\beta}-\frac{1}{4}\chi T_{\left(\chi\right)}^{\mu\nu}-\frac{1}{2}\left(\chi^{\lambda}\nabla_{\lambda}\right)T_{(\chi)}^{\mu\nu} 
\end{equation}
 
when $ T(\chi)=T_{\lambda}^{\lambda}(\chi) $  is the trace of $ T_{(\chi)}^{\mu\nu} $. We see that the stress-energy tensor $ T_{eff}^{\mu\nu} $  is related to original $ T_{(\chi)}^{\mu\nu} $ but it is not equal to it. Where $ T_{eff}^{\mu\nu} $  is the stress-energy momentum tensor that apear in the right side of Einstein equation  $T^{\mu\nu}_{eff}=R^{\mu\nu}-\frac{1}{2}g^{\mu\nu}R$. As we can see, there is a simple linear relation between $  \frac{1}{4}\chi T_{\left(\chi\right)}^{\mu\nu} $, and $ \frac{1}{2}\left(\chi^{\lambda}\nabla_{\lambda}\right)T_{(\chi)}^{\mu\nu} $  in some spaces the last expression could be proportional to $ T_{\left(\chi\right)}^{\mu\nu} $
 if such a tensor is an homogeneous function, by using Euler's homogeneity theorem, which it is satisfied for power law functions of time (in Cosmological solutions). In addition to conservation of $ T_{\left(\chi\right)}^{\mu\nu} $
  the original $ T_{eff}^{\mu\nu} $ also should be conserved. This two conditions are co-exist on the same metric of space-time.

In a context of Cosmology, because of the presence of the term $ \frac{1}{2}\left(\chi^{\lambda}\nabla_{\lambda}\right)T_{(\chi)}^{\mu\nu} $
  we are interested in using the Friedman-Lemaitre-Robertson-Walker (FLRW) metric, with spacial curvature $ k=0 $, wich is the only one that gives the property of proportionality between those two kinds of stress-energy tensors $ T_{\left(\chi\right)}^{\mu\nu} $ and $ T_{eff}^{\mu\nu} $. Otherwise, we get a contradiction between the two conservation of stress-energy momentum tensors. We find this mathematical condition in our universe, that from measurements we sure that there isn't spacial curvature, or if it's exists, it is very small. From our theory, this is a natural consequennce. The only exeption to this rule, would be a very special case, where $a(t)=t$, and k could non zero (like for example, the Milne universe). This will be discussed later. \\ For this metric, with $k=0$, the scale parameter should be $ a(t)=t^{\alpha} $, when $ \alpha $ is the index of the scale parameter: $ \alpha=\frac{1}{2} $
  for radiation, and $ \alpha=\frac{2}{3} $
  for dust (dark matter). With the homothetic vector filed of the metric, using (4) we get: 
\begin{equation}
\chi^{\mu}=\frac{\chi}{4}\left(t,(1-\alpha)\overrightarrow{x}\right)
\end{equation}
 For this space we obtain that: $ \chi_{\mu;\nu}=\frac{\chi}{4}g_{\mu\nu} $ with no anti-symmetrical contribution to the tensor $ \chi_{\mu;\nu} $.
\section{Radiation Like Scalar Field}

Some approches in the context of Two Measure Theories [3] have shown that a scalar field can act as dust (dark matter). Here we will see that in our approach a scalar field can have the behavior of radiation. We choose a stress energy tensor of scalar filed $ \phi $, which doesn't have a trace $ T(\chi)=0 $, therefore we choose:

\begin{equation}
T_{(\chi)}^{\mu\nu}=(g^{\mu\alpha}g^{\nu\beta}-\frac{1}{4}g^{\mu\nu}g^{\alpha\beta})\phi_{,\alpha}\phi_{,\beta}
\end{equation}

From the Cosmological Principle the dependence of the scalar filed is only a function of time $ \phi(t) $. From (7) we get $ T_{\nu(\chi)}^{\mu}=\rho_{(\chi)}diag\left\{ 1,-1/3,-1/3,-1/3\right\}  $ when $ \rho(\chi)=(\frac{d}{dt}\phi)^{2} $. From conservation of $ T_{\left(\chi\right)}^{\mu\nu} $
 we get $ \rho(\chi)\sim a^{-4} $, like radiation. From variation of $ \phi $ we get a covariantly conserved current:

\begin{equation}
j_{(\chi)}^{\mu}=\left(\chi^{\mu;\nu}+\chi^{\nu;\mu}\right)\phi_{,\nu}-\frac{1}{2}\chi\phi^{,\mu}
\end{equation}

The conservation of the current $ j_{;\mu}^{\mu}(\chi)=0 $
  may be a condition for the vector field $ \chi^{\lambda} $. For the ansatz (4) of homothetic Killing vector we get $ j_{(\chi)}^{\mu}=0 $. This obviously satisfies the covariant conservation of (8). From the variation of the metric we get the effective stress-energy momentum tensor,which is the right hand side of the Einstein Tensor, then (5) reduse to the term:

\begin{equation}
T_{eff}^{\mu\nu}=-\frac{\chi}{2}T_{(\chi)}^{\mu\nu}-\frac{1}{2}\left(\chi^{\lambda}\nabla_{\lambda}\right)T_{(\chi)}^{\mu\nu}=-\frac{1}{2}(\chi^{\lambda}T_{(\chi)}^{\mu\nu})_{;\lambda}
\end{equation}

Two Friedman equation's are reduced to one, because of the tracelessness of the stress-energy tensor. Then we take the '00' compoent for the and $\rho(\chi)\sim a^{-4} $, then we get:

\begin{equation}
\rho_{eff}=-\frac{\chi}{2}(1+\frac{1}{2}t\frac{\partial}{\partial t})\rho_{(\chi)}=\chi\frac{\alpha-1}{2}\rho_{(\chi)}
\end{equation}

when Hubble Constant is defined as $ H=\frac{\dot{a}}{a} $
  (for a power law universe $ H=\alpha/t $). That means $ \rho_{eff} $ is proportional to $ \rho(\chi) $. We can get from that $ a\propto t^{1/2} $,because of the relation $ \rho_{eff}\propto H^{2} $, which cames from Einstein equation, for FRLW metric. The constans factor between is with the constant factor between $ \rho_{eff} $ and $ \rho(\chi) $ is $ -\frac{\chi}{4}$. We understand that the physical meaning of this theory is that a scalar filed could act like a radiation field, and could co-exist with the backround radiation in the early universe. The question we would like to search is do this model of scalar filed can predict any existence of dust or dark energy, if there is any potential for this field.

 \section{Electromagnetic Radiation}
  Another possibilty to explore is when $ T_{(\chi)}^{\mu\nu} $ will be like the standard Electromagnetic stress-energy mommentum tensor:
  
 \begin{equation}
  T_{(\chi)}^{\mu\nu}=F^{\mu\alpha}F_{\alpha}^{\nu}-\frac{1}{4}g^{\mu\nu}F_{\alpha\beta}F^{\alpha\beta}
 \end{equation}
  
  where $ F_{\mu\nu}=\partial_{\mu}A_{\nu}-\partial_{\nu}A_{\mu} $. From a variation of $ A_{\mu} $  we get some anti-symmetric tensor, which can be interpreted as a “dielectric” field tensor:
  
  \begin{equation}
D^{\mu\nu}=-\chi^{\mu;\beta}F_{\beta}^{\nu}+\chi^{\nu;\beta}F_{\beta}^{\mu}-\chi^{\beta;\nu}F_{\beta}^{\mu}+\chi^{\beta;\mu}F_{\beta}^{\nu}-\chi F^{\mu\nu}
 \end{equation}

  which has no sources $ (\nabla_{\mu}D^{\mu\nu}=0
   ) $. For the ansatz (4) of homothetic Killing vector we get $ D^{\mu\nu}=0 $
   , like the effective current of the radiation like scalar field. In this case we will get the same identity like (11), and the field in dynamical time will give the same predictions abuot the behavior of the expantion of the universe. Unlike the case of a scalar filed, which is automatically homogeneous and isotropic, now for the $ A_{\nu} $
    field, we obtain that in order to regain homogenity and isotropy, we have to consider a statistical average. 
  
  More general theory, without any scalar fields, may include the traditional action for E.M field and the action from (3). So the total action will be:
  
\begin{equation}
  S=\int d^{4}x[\sqrt{-g}\chi_{\mu;\nu}T_{\left(\chi\right)}^{\mu\nu}+\frac{1}{16\pi G}\sqrt{-g}R-\frac{1}{4}\sqrt{-g}F_{\alpha\beta}F^{\alpha\beta}]
\end{equation}

  Without the last term $ F_{\alpha\beta}F^{\alpha\beta} $
    , Maxwell equations are not obtained, even for spaces where the ansatz (4) of homothetic Killing vector applies, although the energy momentum conservation of $ T_{(\chi)}^{\mu\nu} $
    is obtained. If the space does not satisfies the ansatz (4), then there will be a deviation from Maxwell equations. This deviation will take the following form:
  \begin{equation}
D^{\mu\nu}=-\chi^{\mu;\beta}F_{\beta}^{\nu}+\chi^{\nu;\beta}F_{\beta}^{\mu}-\chi^{\beta;\nu}F_{\beta}^{\mu}+\chi^{\beta;\mu}F_{\beta}^{\nu}-\chi F^{\mu\nu}-F^{\mu\nu}
 \end{equation}
   
    where $\nabla_{\mu}D^{\mu\nu}=0 $. If the space satisfies the ansatz (4) we get the regular Maxwell equations. When the space time does not satisfy the ansatz (4), it aquires non-trivial diaelectric properties, that can exist close to charged massive body or black holes for example, which will change the general behavior of electromanetic field in curved space-time. The stress energy tensor for the case of homothetic Killing vector is:
    \begin{equation}
  T_{eff}^{\mu\nu}=T_{(\chi)}^{\mu\nu}-\frac{1}{2}(\chi^{\lambda}T_{(\chi)}^{\mu\nu})_{;\lambda}
  \end{equation}
  For non-abelian Yang-Mills gauge theories the same effect takes place, where $ F_{\mu\nu}^{a}=\partial_{\mu}A_{\nu}^{a}-\partial_{\nu}A_{\mu}^{a}+gf^{abc}A_{\mu}^{b}A_{\nu}^{c} $ , we will get that similar results analogous to those considered in the abelian case. 
\section{Situation for which $T_{eff}^{\mu\nu}=0$ but $ T_{\left(\chi\right)}^{\mu\nu}\neq0$} 
Looking for example at (15), with respect to cosmological solutions, like FRWM we get $ \rho_{eff}=[1+\chi\frac{\alpha-1}{2}]\rho_{(\chi)}$. In a particular case, the effective gravitational energy is zero $T_{eff}^{\mu\nu}=0$, but it contains radiation which is represented by $ T_{\left(\chi\right)}^{\mu\nu} $ and $\chi=4$. In this case we get that k doesn't have to be zero, in order to have the non trivial homogenity $\frac{1}{2}\left(\chi^{\lambda}\nabla_{\lambda}\right)T_{(\chi)}^{\mu\nu}=nT_{(\chi)}^{\mu\nu} $ which we discused at (5). Taking the case of Milne Universe ($k=-1$ and $a(t)\sim t$), and taking the case of homothetic Killing vector which is of the form $\chi^{\mu}=c(t,0,0,0)$, we obtained that $\rho_{(\chi)}\sim a^{-4}$, and because it's a flat universe, and  $\rho_{(\chi)}\sim t^{-4}$, which is a boundery between inflationary behavior ($\alpha>1$) and non-inflationary ($1>\alpha>0$) behavior.
\section{Discussion, Conclusions and Prospects} 
   
   We have found that a generalization of TMT gives rise to two non trivial energy momentum tensors which are related but not equal in general. Because of the vector field variation, it enforces its covariant conservation of $ T_{\left(\chi\right)}^{\mu\nu} $
     in addition to $ T_{eff}^{\mu\nu} $
    . We have studied this for the case of the vector field as homothetic Killing vector, and using $ T_{\left(\chi\right)}^{\mu\nu} $
     of radiation or radiation like scalar filed Cosmological solutions. In the radiation era of the universe, the $ T_{eff}^{\mu\nu} $
     stress energy tensor could co-exist with the original radiation, which is the source of the Backround Cosmic Radiation. The solution, which could be a model for the early universe, requires no spacial curvature for the universe, because those solutions require homogenity of the dependence on time of the scale factor $ a(t)=t^{\alpha} $
     in order for the $ T_{\left(\chi\right)}^{\mu\nu} $
     and $ T_{eff}^{\mu\nu} $
     to be proportional to each other. This mathematical explanation singles out zero spacial curvature. So it may be an alternative explanation for negligible spacial curvature instead of inflation [4]. In any case, inflation still takes care of other problems, like the horizon problem.
   
   These kind of models could be used in the context of inlation. It will be interesting to add a potential to this scalar filed, and see if one can reproduce an inflationary behavior, also a quantum behavior of this field, and its affect on the first moments of the universe as an inflationary model.
   
   When $ T_{\left(\chi\right)}^{\mu\nu} $
     is a standard stress energy tensor of gauge filed, we obtain similar conclusions. Also this could lead to a modified electromagnetism and similar situations for other gauge theories in curved space time. We would like to find the spherically symmetric charged solutions like charged black hole.
     
     \section{Appendix - identities}
     $$\frac{\partial g^{\alpha\beta}}{\partial g_{\mu\nu}}=-\frac{1}{2}(g^{\alpha\mu}g^{\beta\nu}+g^{\alpha\nu}g^{\beta\mu})$$\\$$\frac{\partial\Gamma_{\lambda\sigma}^{\tau}}{\partial g_{\mu\nu}}=-\frac{1}{2}(g^{\mu\tau}\Gamma_{\lambda\sigma}^{\nu}+g^{\nu\tau}\Gamma_{\lambda\sigma}^{\mu})$$\\$$\frac{\partial\Gamma_{\lambda\alpha}^{\tau}}{\partial g_{\mu\nu,\sigma}}=\frac{1}{4}\left[g^{\mu\tau}\left(\delta_{\alpha}^{\nu}\delta_{\lambda}^{\sigma}+\delta_{\lambda}^{\nu}\delta_{\alpha}^{\sigma}\right)+g^{\mu\nu}\left(\delta_{\alpha}^{\mu}\delta_{\lambda}^{\sigma}+\delta_{\lambda}^{\mu}\delta_{\alpha}^{\sigma}\right)-g^{\tau\sigma}\left(\delta_{\alpha}^{\mu}\delta_{\lambda}^{\nu}+\delta_{\lambda}^{\mu}\delta_{\alpha}^{\nu}\right)\right]$$
     \\$$T_{eff}^{\alpha\beta}=\frac{1}{\sqrt{-g}}\frac{\partial\left(\sqrt{-g}\chi_{\mu;\nu}T_{\left(\chi\right)}^{\mu\nu}\right)}{\partial g_{\alpha\beta}}-\frac{1}{\sqrt{-g}}\frac{\partial}{\partial x^{\sigma}}\frac{\partial\left(\sqrt{-g}\chi_{\mu;\nu}T_{\left(\chi\right)}^{\mu\nu}\right)}{\partial g_{\alpha\beta,\sigma}}$$


\begin{thebibliography}{4}
     \bibitem{1} 
     E.I. Guendelman.
     Mod. Phys. Lett. A14 , 1043 (1999)e-Print: gr-qc/9901017 , E.I. Guendelman and A.B. Kaganovich, Phys. Rev. D53 , 7020 (1996); Phys. Rev. D60 , 065004 (1999) ; F. Gronwald, U. Muench, A. Macias, F. W. Hehl, Phys. Rev. D58 , 084021 (1998), e-Print: gr-qc/9712063. E.I. Guendelman, A.B. Kaganovich, Class. Quantum Grav. 25 , 235015 (2008), e-Print: arXiv:0804.1278 [gr-qc] ; H. Nishino, S. Rajpoot, Mod. Phys. Lett. A21 , 127 (2006), e-Print: hep-th/0404088; E.I. Guendelman, O. Katz, Class. Quantum Grav. 20 , 1715 (2003) e-Print: gr-qc/0211095; E.I. Guendelman, A.B. Kaganovich, Ann. of Phys. 323 , 866 (2008), e-Print: arXiv:0704.1998 [gr-qc] ; E.I. Guendelman, A.B. Kaganovich, Phys. Rev. D75 , 083505 (2007), e-Print: gr-qc/0607111; E.I. Guendelman, A.B. Kaganovich, Int. J. Mod. Phys. A21 , 4373 (2006), e-Print: gr-qc/0603070 ; E.I. Guendelman, A.B. Kaganovich , Int. J. Mod. Phys. A17 , 417 (2002), e-Print: hep-th/0110040, Eduardo Guendelman, Ramón Herrera, Pedro Labrana, Emil Nissimov, Svetlana Pacheva , Gen.Rel.Grav. 47 (2015) no.2, 10 DOI: 10.1007/s10714-015-1852-1 e-Print: arXiv:1408.5344 [gr-qc] and references therein. 
      
     \bibitem{2} 
     E.I. Guendelman. Annalen der Physik, 322(10):891–921, Int.J.Mod.Phys. A25 (2010) 4081-4099 DOI: 10.1142/S0217751X10050317 e-Print: arXiv:0911.0178 [gr-qc]. 
           
     \bibitem{3} 
     Eduardo Guendelman, Emil Nissimov, Svetlana Pacheva. Conference: C15-06-15.6 e-Print: arXiv:1512.01395 [gr-qc] .Eduardo Guendelman, Emil Nissimov, Svetlana Pacheva. Published in Eur.Phys.J. C76 (2016) no.2, 90 DOI: 10.1140/epjc/s10052-016-3938-7 e-Print: arXiv:1511.07071 [gr-qc].
      Eduardo Guendelman, Emil Nissimov, Svetlana Pacheva. Published in Eur.Phys.J. C75 (2015) no.10, 472 DOI: 10.1140/epjc/s10052-015-3699-8 e-Print: arXiv:1508.02008 [gr-qc].
       
     \bibitem{4} 
          E.W. Kolb and M.S. Turner 
         “The Early Universe” , Addison Wesley (1990); A. Linde, “Particle Physics and Inflationary Cosmology” , Harwood, Chur, Switzerland (1990); A. Guth, “The Inflationary Universe” , Vintage, Random House (1998); A.R. Liddle and D.H. Lyth, “Cosmological Inflation and Large-Scale Structure” , Cambridge Univ. Press (2000); S. Dodelson, “Modern Cosmology” , Acad. Press (2003); S. Weinberg, “Cosmology” , Oxford Univ. Press (2008).
     
     \end{thebibliography}
\end{document}